\documentclass[aps,twocolumn,superscriptaddress,showpacs,showkeys]{revtex4}
\usepackage{graphics,graphicx,dcolumn,bm,fleqn,epic,eepic,float,epsfig}
\usepackage{amssymb,amsmath,multirow,rotate,color,float}
\usepackage{epstopdf}
\usepackage{times}
\usepackage{color}
\usepackage{soul}                    
\definecolor{red}{rgb}{1,0,0}
\definecolor{green}{rgb}{0,1,0}
\definecolor{blue}{rgb}{0,0,1}

\begin{document}

\title{Deposition of general ellipsoidal particles}

\author{Reza M.~Baram}
\email{reza@cii.fc.ul.pt}
\affiliation{Center for Theoretical and Computational Physics, 
             University of Lisbon,
             Av.~Prof.~Gama Pinto 2, 1649-003 Lisboa, Portugal}

\author{Pedro G.~Lind}
\email{plind@cii.fc.ul.pt}
\affiliation{Center for Theoretical and Computational Physics, 
             University of Lisbon,
             Av.~Prof.~Gama Pinto 2, 1649-003 Lisboa, Portugal}
\affiliation{Department of Physics, Faculty of Sciences of the
             University of Lisbon, 1649-003 Lisboa, Portugal} 

\date{\today}

\begin{abstract}
We present a systematic overview of granular deposits composed of
ellipsoidal particles with different particle shapes and size 
polydispersities.  We study the density and anisotropy of such 
deposits as functions of size polydispersity and two shape parameters 
that fully describe the shape of a general ellipsoid.
Our results show that, while shape influences significantly the
macroscopic properties of the deposits, polydispersity plays
apparently a secondary role. The density attains a maximum for a 
particular family of non-symmetrical ellipsoids, larger than the 
density observed for prolate or oblate ellipsoids.
As for anisotropy measures, the contact forces show 
are increasingly preferred along the vertical direction as 
the shape of the particles deviates for a sphere.  
The deposits are constructed by means of an 
efficient molecular dynamics method, where the contact forces are efficiently
and accurately computed.  The main results are discussed in the light of 
applications for porous media models and sedimentation processes.

\end{abstract}



\maketitle


\section{Introduction}

The granular character of many processes in nature has motivated
the study of granular materials and in particular granular packings
since Kepler's time\cite{torquato2010}. One of such processes is
the deposition of particles subject to gravity\cite{kadau2011},
which underlies the formation of sandstones, ceramics and in 
general, the emergence of porous materials.  
Modelling porous media\cite{joshi2010,biswal2009}
in a realistic way is important for instance
to understand permeability of soils for petroleum engineering or
in material sciences.

However, a model for grain deposition should incorporate 
two ingredients not always easy to combine. 
First, empirical studies\cite{sukumaran2009} show that the shape of grains 
significantly deviates from the sphere with a non-negligible polydispersity 
in their sizes.
Until recently most computational studies dealt with round particles
for simplicity, in solving specific problems such as 
obtaining the optimal packing\cite{schneider2009,reis2011}, constructing 
space-filling configurations\cite{reza2004,reza2005,lind2008}
or studying elongated shapes\cite{hidalgo2011}.
Some progress was recently made when addressing the influence of 
particle shape on the macroscopic observables of the agglomerate, 
focusing on ellipsoidal particles\cite{donev04b,donev07} 
and even in more general shapes\cite{delaney2010,jiao2008},
without accounting for the size polydispersity. 

Second, deposition processes are fundamentally sequential and therefore
should be modelled through sequential procedures.
The procedure introduced by Donev 
{\em et al}\cite{donev04b,torquato2010} for jammed configurations,
captures some of the general features observed in granular deposits, 
such as their density and average coordination number, although it
is based on a non-sequential procedure. However,
deposition under gravity does not generally lead to a jammed state.
Recently, some sequential procedures were already
carried out in two-dimensional deposits of mono-disperse elongated 
particles\cite{kadau2011,hidalgo2011}.

In this paper we focus on three-dimensional polydisperse deposits of 
general ellipsoids and study how macroscopic properties of 
such deposits depend on the shape and size polydispersity of 
particles. 
We show that the density and coordination number of deposits behaves
similarly to what is observed in jammed systems. However, in contrast
with randomly constructed jammed systems, deposits become strongly 
anisotropic when the shape of the particles deviate from the sphere.
Furthermore, we find that the size polydispersity plays a minor role
when compared to the shape.

To this end, we develop an efficient algorithm for the calculation of the 
interactions forces at all contact points within a three-dimensional 
deposits of polydisperse ellipsoids and use it in a 
Molecular Dynamics\cite{granmatt,mdsimul} method to perform our simulations. 
Further, we introduce two shape parameters which describe any
ellipsoid, enabling a systematic study of the role of particle
shape in granular deposits.

The outline of the paper is as follows. 
In Sec.~\ref{sec:model} we describe the algorithm for 
calculating the contact forces between colliding ellipsoids with arbitrary 
size and shape as well as the simulation setup.
In Sec.~\ref{sec:controlparam} the control parameters describing
the polydispersity in the size and the shape of a general
ellipsoid are introduced.
In Sec.~\ref{sec:macroprop} the dependency of macroscopic observables
taken from the entire deposit are studied as functions of the control 
parameters.
We focus on the density, coordination number and anisotropy of 
the deposits.
Discussion and conclusions are given in Sec.~\ref{sec:conclusions}.

\section{Modeling the contact forces between ellipsoidal particles}
\label{sec:model}

When dealing with large systems, calculating the inter-particle forces 
is by far the most time consuming part of the computations. This is due to 
the fact that, in general, every particle can interact with every other one. 
In granular materials, however, the inter-particle forces are short range 
making techniques such as verlet list and linked-cell algorithm 
indispensable in reducing the computation time. These technique are employed 
to eliminate the redundant calculations for pairs which are 
too far apart to have any interaction\cite{granmatt}.

\begin{figure}[t] 
\begin{center}
\includegraphics*[width=0.45\textwidth,angle=0]{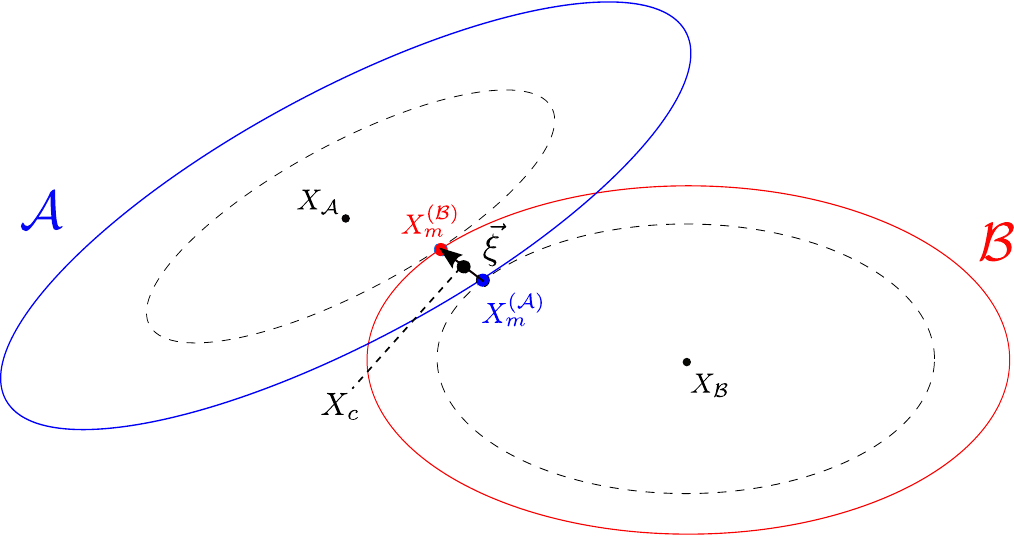}
\caption{\protect
         (Color online) 
         Sketch of the contact vector $\vec\xi$ 
         used to define the normal repulsive 
	 force between two overlapping ellipsoids 
         $\mathcal{A}$ and $\mathcal{B}$.
         The contact point is chosed to be $X_c$ 
         after properly deriving points $X_m^{(\mathcal{A})}$
         and $X_m^{(\mathcal{B})}$ at the surfaces of 
         the ellipsoids (see text). The full procedure is
         sketched in Fig.~\ref{fig:diagram}.}
\label{fig:contact}
\end{center}
\end{figure}

In time-driven Molecular Dynamics (MD) methods of granular materials 
the system is evolved for one time step allowing the particles to overlap. 
Then normal repulsion forces on the particles are calculated as a function 
of the amount of overlapping, which is defined usually by a line segment. 
Here we refer to this line segment as the contact vector 
(see Fig.~\ref{fig:contact}).

The contact vector between two ellipsoids can be defined in more than one way.
The common method, introduced by Perram and co-workers\cite{perram85}
and developed by others\cite{perram96,donev07,lind09}, is based on finding 
two points, one on the surface of each ellipsoid by minimizing a 
potential through a variational problem with one geometric constraint.
The two obtained points can be used to define the contact vector.
However, the contact vector defined this way is not necessarily normal to 
the surface of the particles, particularly when particles are
strongly asymmetrical, leading to non-physical motions of the particles.

An alternative method consists in finding two points on the surfaces of 
the ellipsoids at which the corresponding surface normals point exactly 
in opposite directions. 
Consequently, the vector connecting these two points is normal to the 
surfaces of both ellipsoids and can be used as a well-defined contact 
vector. This method is, however, much more computationally 
expensive than the former one. See Ref.~\cite{lin95} for more details. 

To retain the computational efficiency, we use the former method and 
introduce an additional step for correcting the direction
of the contact vector.

The geometric potential of the ellipsoid $\mathcal{A}$ is defined as:
\begin{equation}
   f_{\mathcal{A}}(X)= (X-X_{\mathcal{A}})^T {\bf A} (X-X_{\mathcal{A}}),
\end{equation}
where $X$ and $X_{\mathcal{A}}$ are the Cartesian coordinates of a point 
on the surface and the centroid of the ellipsoid, respectively. 
The equation $f_{\mathcal{A}}(X)=1$ describes the surface of ellipsoid $\mathcal{A}$.
Considering a pair of ellipsoids $\mathcal A$ and $\mathcal B$, 
two minimum points $X_m^{(\mathcal{A})}$ and $X_m^{(\mathcal{B})}$ 
are found on the surfaces of $\mathcal{A}$ and $\mathcal B$
by minimizing  $f_{\mathcal{B}}$ and $f_{\mathcal A}$, respectively.
These points are shown schematically in Fig.\ref{fig:contact}. 
To obtain the minima of $f_{\mathcal{A}}(X)$  subject to the constrain of 
being on the surface of ellipsoid $\mathcal{B}$ we minimize the following 
auxiliary potential:
\begin{equation}
   f(X)= (X-X_{\mathcal{A}})^T {\bf A} (X-X_{\mathcal{A}}) + 
          \lambda \left [ (X-X_{\mathcal{B}})^T {\bf B} (X-X_{\mathcal{B}})-1\right],
\end{equation}
where $\lambda$ is the Lagrange multiplier associated with the constrain. 
Setting to zero the gradient of this function with respect to $X$, 
\begin{equation}
   \label{eq:lagrange}
   \nabla f(X) = {\bf A} (X_m-X_{\mathcal{A}}) + 
    \lambda {\bf B} (X_m-X_{\mathcal{B}})=0,
\end{equation}
we obtain an equation for {\em optimum} points $X_m$ for $\mathcal{A}$:
\begin{equation}
    \label{eq:min}
    X_m
          =({\bf A}+\lambda {\bf B})^{-1} ({\bf A} X_{\mathcal{A}} + 
                 \lambda {\bf B} X_{\mathcal{B}}).
\end{equation}

To derive an equation for $\lambda$ we left-multiply Eq.~(\ref{eq:lagrange}) 
by $ (X_m-X_{\mathcal{B}})^T $ and use the fact that $X_m$ lies on $\mathcal{B}$ 
and that the gradients of the potential functions of the ellipsoids
 at {\em minimum} point are in opposite directions: 
\begin{equation}
    \label{eq:lambda}
    \lambda= \left |  (X_m-X_{\mathcal{B}})^T {\bf A} (X_m-X_{\mathcal{A}})  
             \right |.
\end{equation}
Equations (\ref{eq:min}) and (\ref{eq:lambda}) can be solved 
iteratively to find $X^{(\mathcal A)}_m$ up to the desired precision. 
The second point  $X^{(\mathcal B)}_m$ can be found similarly. 

The contact point $X_c$ where a contact force is exerted on both ellipsoids 
is defined as the midpoint between $X_m^{(\mathcal{A})}$ and $X_m^{(\mathcal{B})}$.
The contact vector for ellipsoid $\mathcal{A}$ is defined as 
${\vec\xi}_{\mathcal{A}}=X_m^{(\mathcal{A})}-X_m^{(\mathcal{B})}$ 
and for ellipsoid $\mathcal{B}$, 
${\vec\xi}_{\mathcal{B}}=-{\vec\xi}_{\mathcal{A}}$ (see Fig.~\ref{fig:contact}).

Although Eq.~(\ref{eq:min}) involves inversion of a $3\times 3$-matrix, 
finding the minima can be very efficient by choosing the initial points 
properly.  Good approximations for minima are the minima from the last time step or, 
if the ellipsoids were disjoint in the last time step, the closest points 
on the surfaces of the ellipsoids which are known from collision detection procedure 
described later in this section. These points are then used as initial points, 
reducing the number of iterations significantly. 

The overlapping vector calculated in this way, however, is not necessarily 
normal to the ellipsoids at the contact point.
In fact, the normal directions to $\mathcal{A}$ and $\mathcal{B}$
at points $X_m^{(\mathcal{A})}$ and $X_m^{(\mathcal{B})}$ deviate significantly
from each other when the particles are strongly aspherical.
To overcome this problem we introduce 
an additional step for correcting the contact vector, as described next. 

If the calculated contact vectors deviate more than a given amount from the 
gradients of the ellipsoid's potential at the contact point $X_c$, a new 
direction is calculated by averaging the gradients at $X_c$.
Then, two new points are calculated from the intersection of the line along 
this direction passing through the contact point. 
These new points are used to define a new contact vector.
Although it is not generally possible to define a vector which is along 
the gradients of both ellipsoids at the contact point,
this approximation gives satisfactory results in terms of energy 
conservation.  

Having described the procedure for calculating the contact of two 
colliding ellipsoids, we next need an efficient way to determine when 
two previously disjoint ellipsoids collide. 
To this end, we use the technique proposed by Wang {\em et al}\cite{wang04} 
which is described briefly as follows.

In homogeneous coordinates the equation for the surface of a general 
ellipsoid is given by
\begin{equation}
   \mathcal{A}: X^T {\bf A} X=0,
\end{equation}
where $X$ is any point on the surface of the ellipsoid and 
${\bf A}$ is the $4\times4$-matrix of the affine 
transformation which transforms the unit sphere centered at the origin into 
the ellipsoid. 
\begin{figure}[floatfix]
\begin{center}
\includegraphics[width=0.48\textwidth,angle=0]{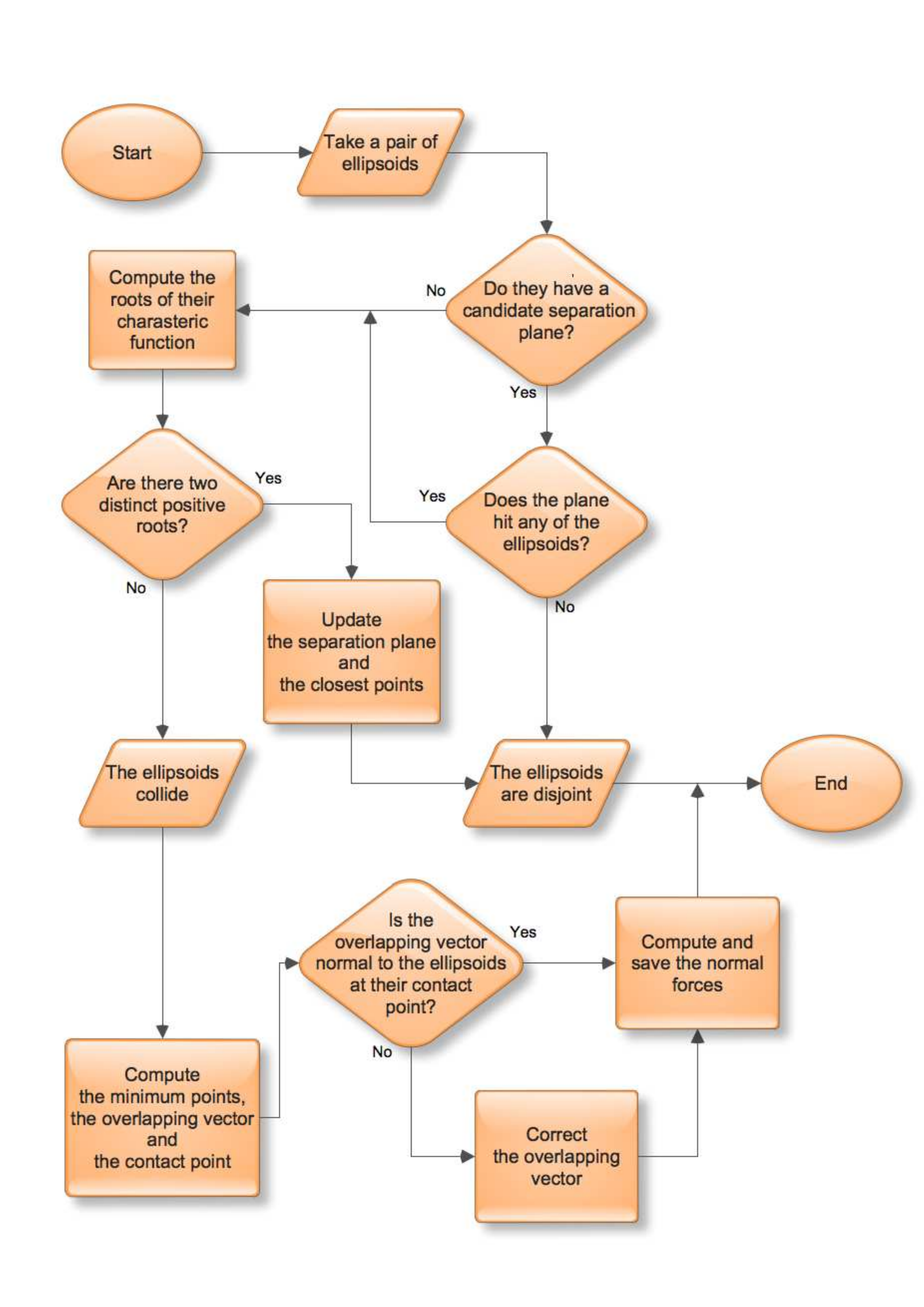}
\caption{\protect
         (Color online)
         Flow-chart of the algorithm for detection and calculation
         of the contact points between two arbitrary ellipsoids.}
\label{fig:diagram}
\end{center}
\end{figure}

Given two ellipsoids $\mathcal{A}: X^T {\bf A} X=0$ and 
$\mathcal{B}: X^T {\bf B} X=0$, their characteristic equation is 
defined as $f(\nu)=\det(\nu {\bf A} +{\bf B})$, whose roots are 
the eigenvalues of matrix $-\mathbf{A}^{-1}\mathbf{B}$. Furthermore,
it can be shown \cite{wang04} that
(i) $f(\nu)=0$ has at least two negative roots,
(ii) $f(\nu)=0$ has two distinct positive roots if and only if the 
two ellipsoids are disjoint,
(iii) $f(\nu)=0$ has a positive double root if and only if the two 
ellipsoids are externally touching.
Therefore, by examining the eigenvalues of $-{\bf A}^{-1}{\bf B}$ it 
can be determined if two ellipsoids collide.

When the ellipsoids are disjoint, the corresponding eigenvectors
give the coordinates of the four vertices $V_i$, $i=1,\dots,4$, of 
a tetrahedron which is self-polar for both ellipsoids $\mathcal{A}$ 
and $\mathcal{B}$\cite{semple}, 
i.e.~$V_i^T\mathbf{A}V_j=V_i^T\mathbf{B}V_j=0, \forall i\neq j$. 
Two of the vertices lie outside both ellipsoids (external vertices), 
while the other two are contained inside the ellipsoids, one in each 
(internal vertices).
In this case, a set of separation planes can be defined using three 
points exterior to both ellipsoids, two of them being the external 
vertices and the other a point on the line segment connecting the 
internal vertices.  
We choose the third point as the midpoint between the surfaces of 
two ellipsoids on this line segment.

Having the separation plane between the two disjoint ellipsoids,
in the subsequent time step we check if the ellipsoids intersect that
plane. 
If not, the ellipsoids are still disjoint and the procedure stops. 
This helps to eliminate the need for checking for collision between 
the ellipsoids which is more computationally expensive. 
Only when any of the ellipsoids intersects the separation plane it is 
necessary to check again for the collision between the ellipsoids.
If the ellipsoids turn out to be still disjoint their separation plane and 
closest points are updated to be used in the next time step. If the 
ellipsoids do collide the contact vector is calculated and saved to be 
used for the calculation of the contact force. 

Figure \ref{fig:diagram} shows the flow-chart of the complete procedure 
which is performed for all potentially overlapping pairs of ellipsoids.
It is worth noting that, 
to further speed up the procedure, only the pairs whose
spherical envelopes intersect are evaluated.

The contact force can be decomposed to normal and tangential forces. 
The tangential forces such as static and dynamical frictions are generally 
derived from the normal force. 
For calculating the magnitude of the normal force we use Hertzian 
model\cite{granmatt}:
\begin{equation}
   \label{eq:hertz}
   F_n\propto  \xi^{3/2}+K\sqrt{\xi}\frac{d\xi}{dt},
\end{equation}
with $\xi=\vert\vec{\xi}\vert$ being the length of the contact vector
between the particles and $K$ being the dissipative constant depending 
on the material viscosity \cite{brilliantov}.  
Since we are only concerned with the properties of the deposit
at rest, we choose $K$ sufficiently large for the deposit to relax 
rapidly. This relaxation is enhanced by including a tangential
force $\vec{F}_t=\mu F_n\vec{v}_t$, $\mu$ being the dynamic
friction coefficient and $\vec{v}_t$ the relative tangent velocity
of the particles.

Finally, the procedure described above is used in a MD simulation
together with standard methods, namely a prediction-correction 
integrator for solving the equations of motions, quaternions for 
describing the orientation of the particles and linked-cell algorithm 
to eliminate unnecessary collision detections. 
For details of the MD methods see e.g.~Ref.~\cite{granmatt} and references therein.

\section{Control parameters for size and shape}
\label{sec:controlparam}

The size and shape of the particles are the main parameters whose 
effects on the properties of the system are to be studied. 
In the following we explain how these are defined and chosen 
when preparing the samples.
\begin{figure}
\begin{center}
\includegraphics*[width=0.45\textwidth,angle=0]{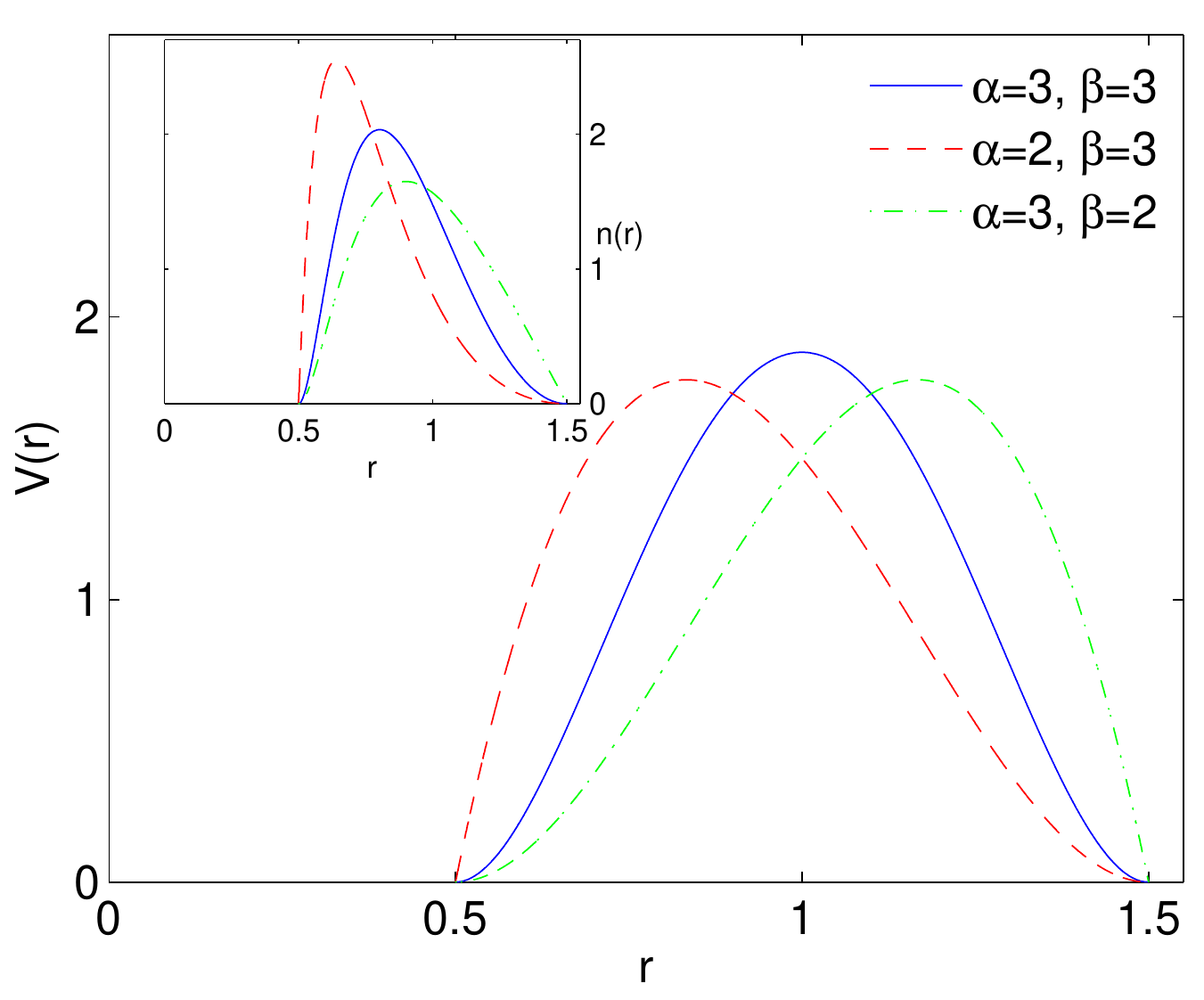}
\caption{\protect
            Volume distributions $V(r)$ for different values of $\alpha$ 
            and $\beta$ as defined in Eq.~(\ref{VV}). 
            The inset shows the corresponding size distributions $n(r)$
            (see text).}
\label{fig:vr-nr}
\end{center}
\end{figure}

We define the size of an ellipsoidal particle as
the radius of the sphere with the 
same volume, i.e.~$r= (abc)^{\tfrac{1}{3}}$ with $a$, $b$ and $c$ being 
the three semi-axis radii of the ellipsoid.  
We refer to a sample as mono-sized or monodisperse in size when all its 
constituting particles have the same volume, i.e.~the same size $r$. 
To introduce size polydispersity we adopt the approach by Voivret 
{\em et al}\cite{voivret2007}: 
instead of directly choosing the size distribution $n(r)$ of the particles, 
we consider the distribution of the total volume 
$V(r)=\frac{4}{3}\pi n(r) r^3$ of all particles with size $r$. 
\begin{figure}
\begin{center}
\includegraphics*[width=0.45\textwidth,angle=0]{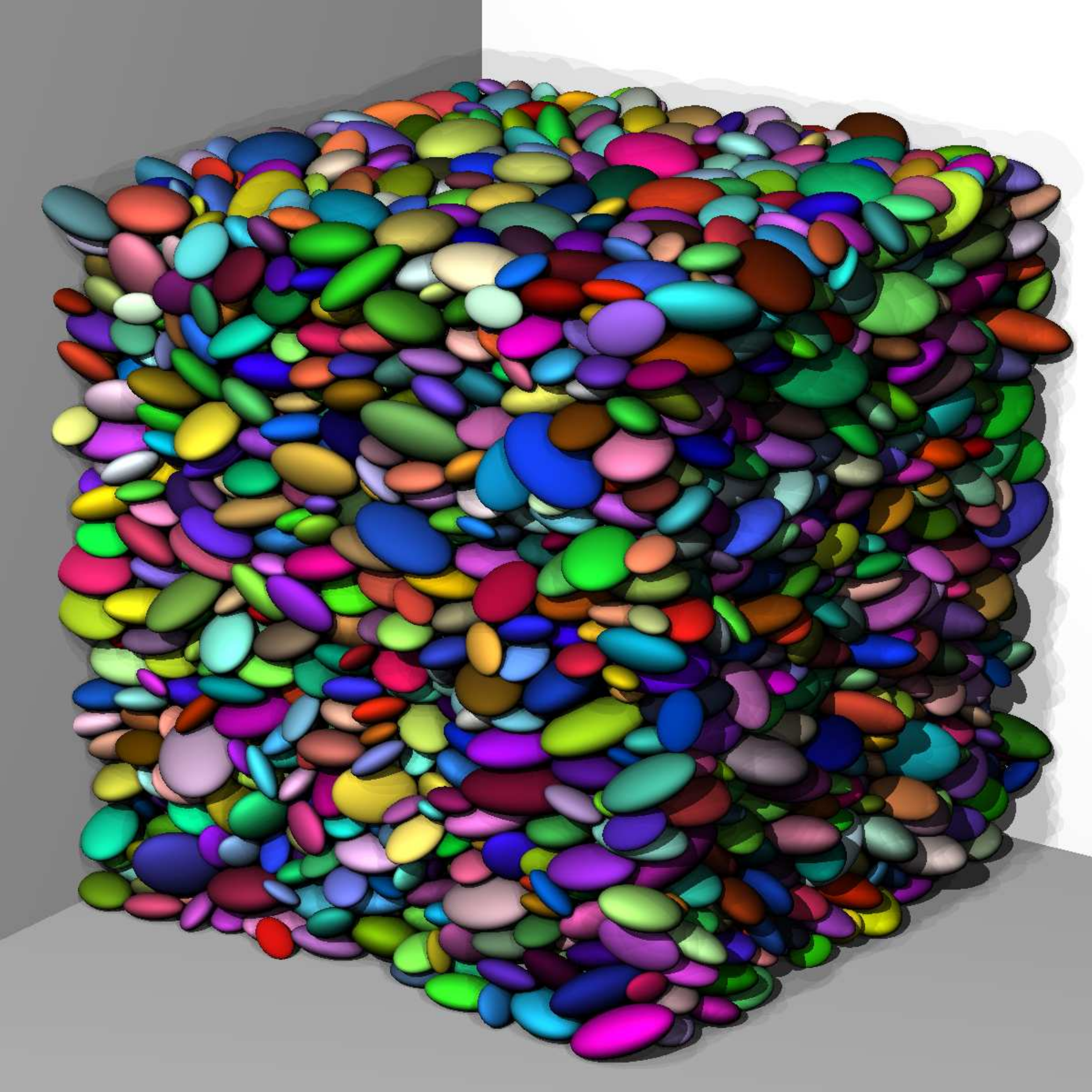}
\caption{\protect
            (Color online)
            A deposit consisting of $N\sim 4200$ ellipsoids 
            with shape parameters $\eta=1.4$ and $\zeta=1.8$ 
            a polydispersity of $\delta_r=0.6$, that is, $r_{min}=0.016$ and 
            $r_{max}=0.064$ (see text).
            Colors are arbitrarily chosen for better visualization.}
\label{fig:deposit}
\end{center}
\end{figure}

We chose $V(r)$ as
\begin{equation}
V(r)=
\left\{
\begin{array}{cc}
C\frac{(r-r_{min})^{\alpha-1}}{(r_{max}-r)^{1-\beta}} & 
          \text{if\ } r_{min}\le r \le r_{max}\\
0 & \text{otherwise,}
\end{array}
\right.
\label{VV}
\end{equation}
where $\alpha$ and $\beta$ determine the shape of the distribution and 
$C$ is the normalization factor. 
Figure \ref{fig:vr-nr} shows $V(r)$ for different values of $\alpha$ 
and $\beta$ and the corresponding size distributions $n(r)$ (inset).  
For $\alpha=\beta$ this function takes a symmetric shape with a peak 
at $(r_{min}+r_{max})/2$. 
For the samples studied in this work we have chosen $\alpha=\beta=3$ 
which corresponds to a distribution (solid line in Fig.~\ref{fig:vr-nr})
very close to a truncated Gaussian.

The width of the distribution is controlled by the polydispersity 
parameter $\delta_r$ which is defined as \cite{voivret2007}:
\begin{equation}
\delta_r=\frac{r_{max}-r_{min}}{r_{max}+r_{min}}.
\end{equation}
For $\delta_r=0$ the particles are monodisperse while $\delta=1$ 
corresponds to infinite polydispersity.

The shape of an ellipsoid is characterized by two parameters
here defined as $\eta=a/b>1$ and $\zeta=b/c>1$, with $a \ge b \ge c$.
For prolate ($a = b < c$) and oblate ($a = b > c$) ellipsoids, the shape 
can be fully characterized by the aspect ratio $a/c=\eta\zeta$. 
The particular case $\eta=\zeta=1$ corresponds to the sphere.
These shape parameters together with the size $r>0$
fully specify any ellipsoid.

Similar to the size polydispersity, two additional parameters 
$\delta_\eta$ and $\delta_\zeta$ can be similarly defined 
for the polydispersities in the shape of the particles. 
In this study we consider systems of mono-shaped particles,
describing each deposit by $\eta$, $\zeta$ and size polydispersity 
$\delta_r$. 

In this work, all the deposits are generated by releasing particles 
with randomly chosen positions and orientations and letting them fall 
in an open box, under gravity along 
the vertical direction.
The box is limited from below by a rigid wall (ground) and is periodic 
in $x$ and $y$ directions.
In order to reduce the boundary effects the first few bottom
layers of particles are not considered in our analysis.
Figure \ref{fig:deposit} shows a deposit of 
ellipsoidal particles with $\eta=1.4$, $\zeta=1.8$ and 
$\delta_r=0.6$.
\begin{figure}
\begin{center}
\includegraphics[width=0.45\textwidth,angle=0]{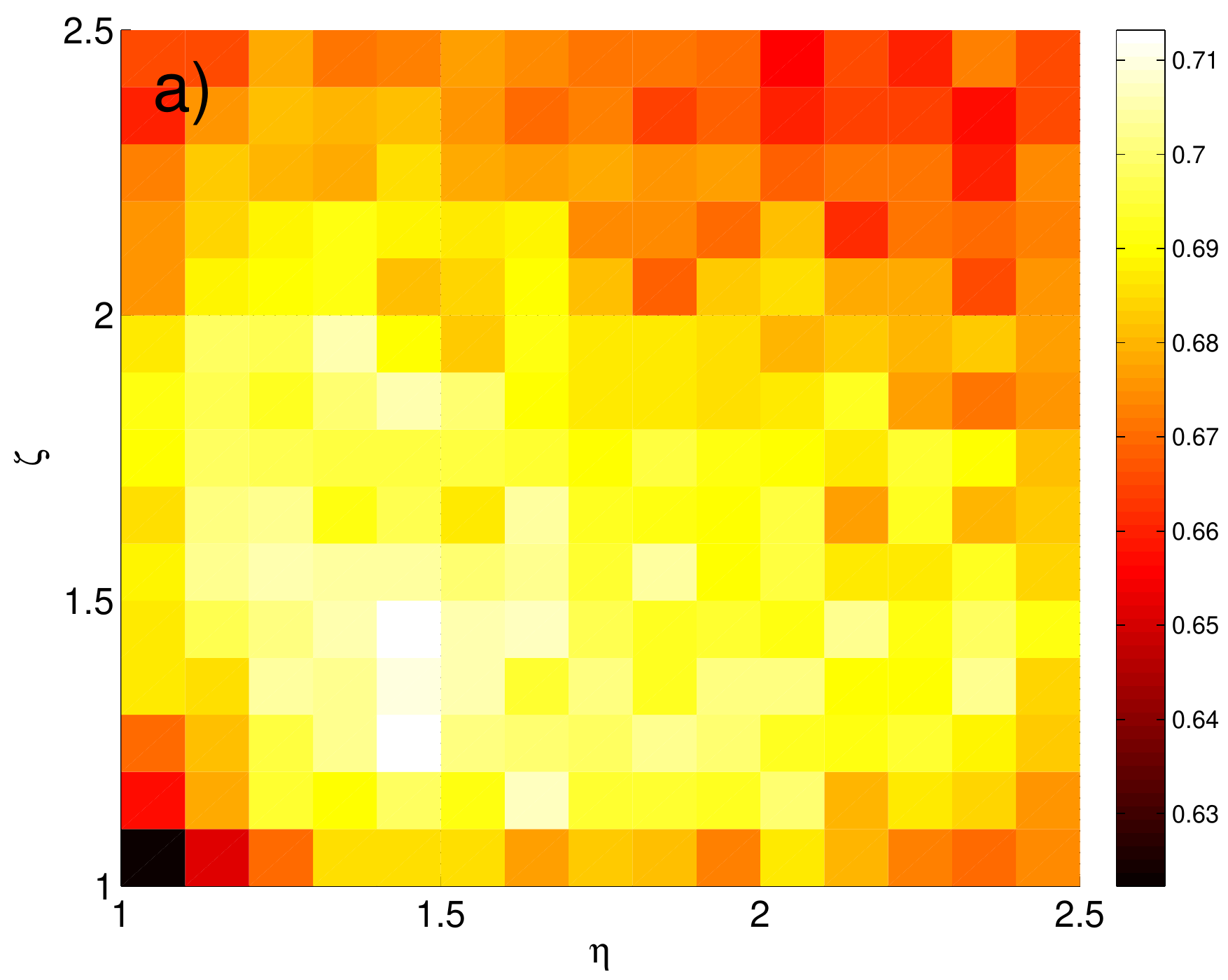}
\includegraphics[width=0.45\textwidth,angle=0]{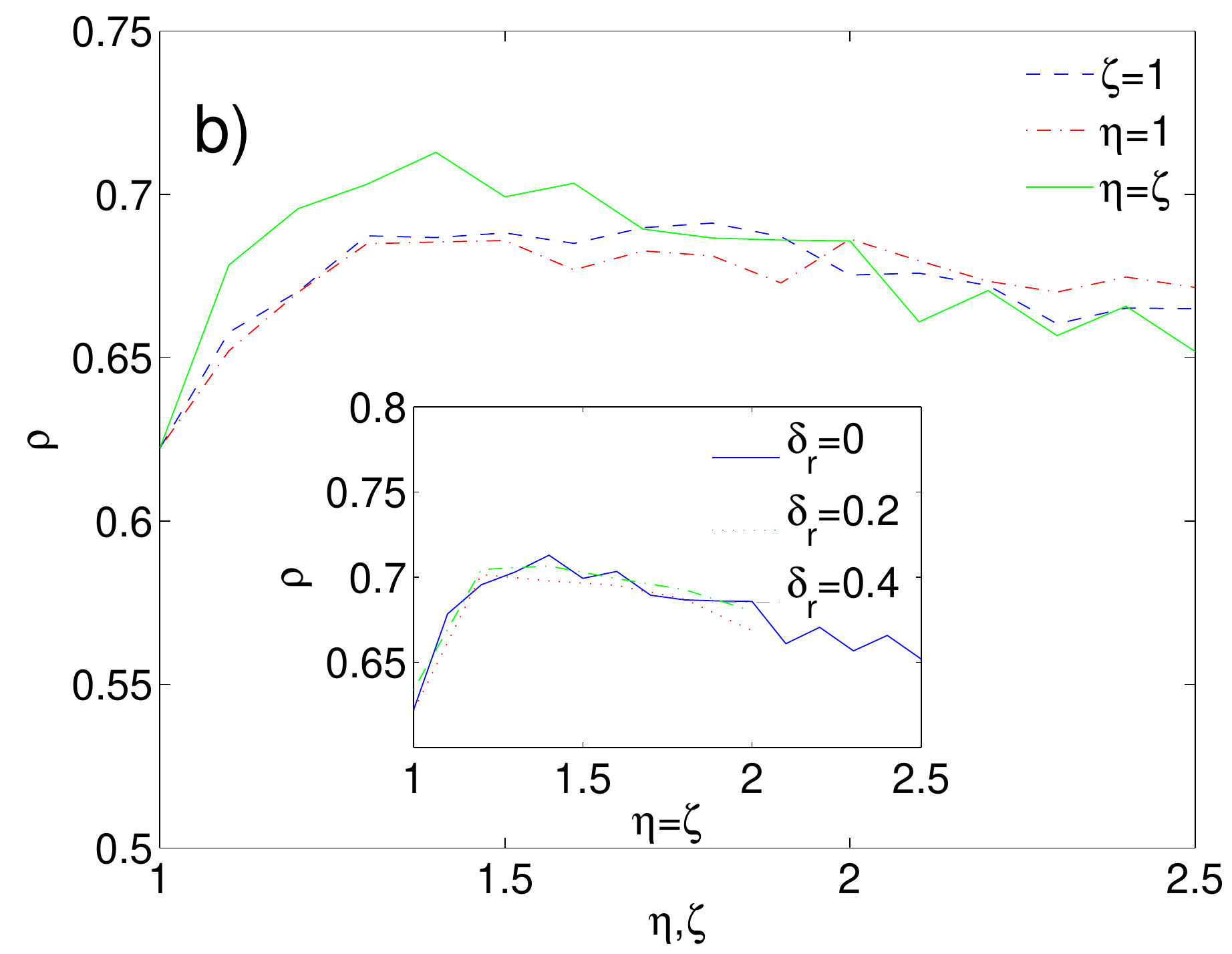}
\caption{\protect
         (Color online)
         {\bf (a)} Density $\rho$ of monodisperse deposits as a 
                   function of the shape parameters $\eta$ and $\zeta$. 
                   The density is approximately 
                   symmetric with respect to the diagonal $\eta=\zeta$
                   where it attains a maximum.
         {\bf (b)} The density as a function of shape along the diagonal 
                   $\eta=\zeta$ (solid line) together with the case of prolate
                   ellipsoids ($\zeta=1$, dashed line) and of oblate 
                   ellipsoids ($\eta=1$, dashed-dotted line). 
                   In the inset the density is plotted as a function of
                   $\eta=\zeta$ for different polydispersities $\delta_r$.
         	   $N\sim 3000-4500$ depending on the amount of the size polydispersity.}
\label{fig:density}
\end{center}
\end{figure}
\begin{figure}
\begin{center}
\includegraphics*[width=0.45\textwidth,angle=0]{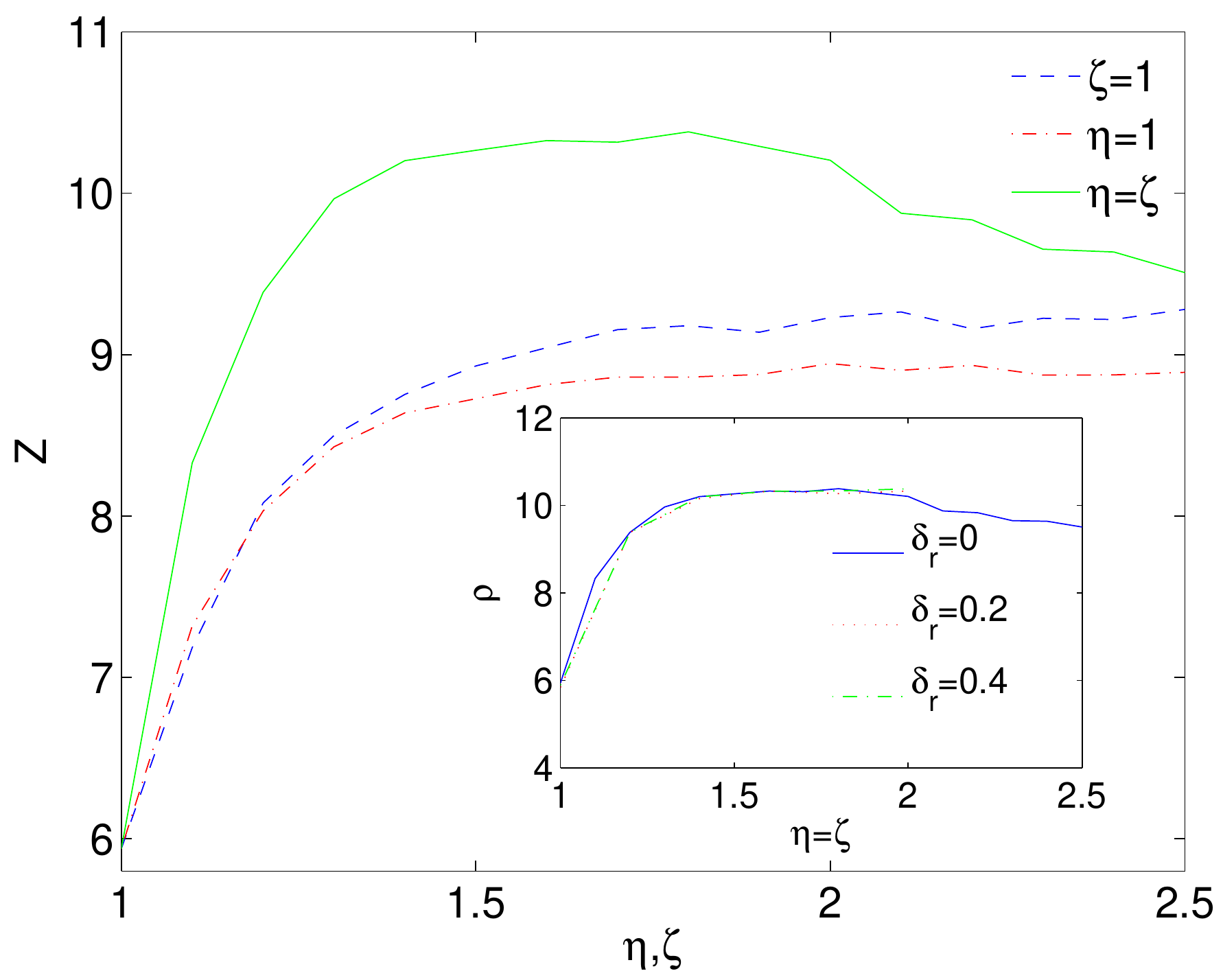}
\caption{\protect
	The average coordination number as a function of the 
        shape parameters $\eta$ and $\zeta$. Here $N\sim 3000$.	}
\label{fig:coordnum}
\end{center}
\end{figure}

\section{Macroscopic properties of the deposit}
\label{sec:macroprop}

The macroscopic properties considered in our study are
the packing density, the average coordination number 
and the anisotropy.

\subsection{Packing density and coordination number}
\label{sec:density}

Figure \ref{fig:density} shows the packing density 
in deposits of $N\sim 3000$ mono-disperse
ellipsoids as a function of the shape. 
The immediate realization is that the density is to a large extend symmetric 
with respect to the axis $\eta=\zeta$.

When the shape deviates from the spherical shape,
i.e.~when $\eta>1$ or $\zeta> 1$,
the density first increases rapidly, then it attains a maximum and finally 
decreases slowly. 
From Fig.~\ref{fig:density}a one sees that typically, along lines of
constant $\eta$ (or constant $\zeta$) the maximum is observed around 
$\eta=\zeta$.
Further, along $\eta=\zeta$ a global maximum
is attained at $\eta=\zeta\approx 1.4$, as can be observed from
Fig.~\ref{fig:density}b (solid line).
This absolute maximum curiously approximates the value of the golden 
ratio $(1+\sqrt{5})/2$, which from the definition of $\eta$ and $\zeta$
corresponds to when $a=b+c$.  Whether such an observation is merely a coincidence or not needs 
further investigations which is out of the scope of our manuscript.

These results are qualitatively similar to those obtained in 
Refs.~\cite{donev07,delaney2010} where a slightly higher 
density is observed,
due to their method for generating the packings:
instead of being generated via deposition, 
their samples are prepared in order to be in jammed state 
using a generalized form of Lubachevsky-Stillinger 
algorithm \cite{lubachevsky90,lubachevsky91}.

The effect of polydispersity is examined by comparing the result for three 
different values of $\delta_r$ as shown in the inset of 
Fig.~\ref{fig:density}b. The results 
indicate that the density seems to be
insensitive to low and moderate size polydispersities. 

Figure \ref{fig:coordnum} shows the average coordination 
number $Z$ as a function of the shape of the particles. 
For spheroidal particles, $Z$ increases gradually with 
the aspect ratios toward a maximum.
For a general ellipsoid the coordination number 
attains a maximum higher than the maximum observed for 
spheroids ($\eta=1$ or $\zeta=1$), 
but decreases to values comparable to those of spheroids for larger
$\eta$ and $\zeta$.  This can be understood considering the fact that
for $\eta=\zeta \gg 1$, one of the semi-axis radii is
significantly larger than the other two and therefore the particle 
can be regarded as a prolate ellipsoid.

\subsection{Anisotropy}
\label{sec:anisotropy}

To study the anisotropy of the deposits we 
investigate the orientational order of 
the contact vectors and the principal directions of the ellipsoids 
by calculating their angular distributions. 
In general two angles are needed to specify the orientation of a vector in 
three dimensional space, namely one azimuthal angle and one angle with the 
vertical direction.
Our deposits are symmetric by construction in azimuthal direction.
This is also observed in our results (not shown here for the sake of brevity). 
Therefore, we will only consider the angle with the positive vertical 
direction.

\begin{figure}[t]
\begin{center}
\includegraphics[width=0.45\textwidth,angle=0]{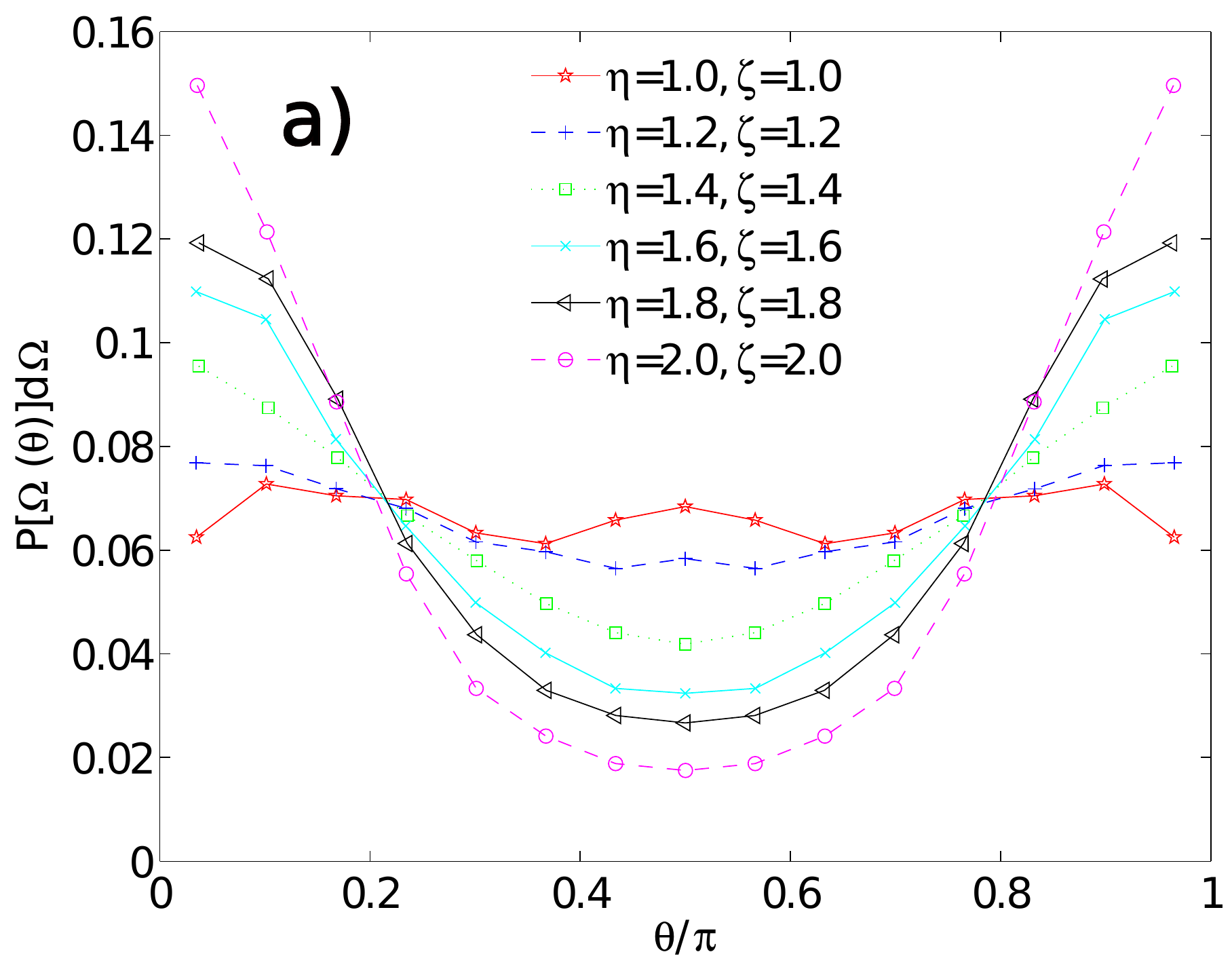}
\includegraphics[width=0.45\textwidth,angle=0]{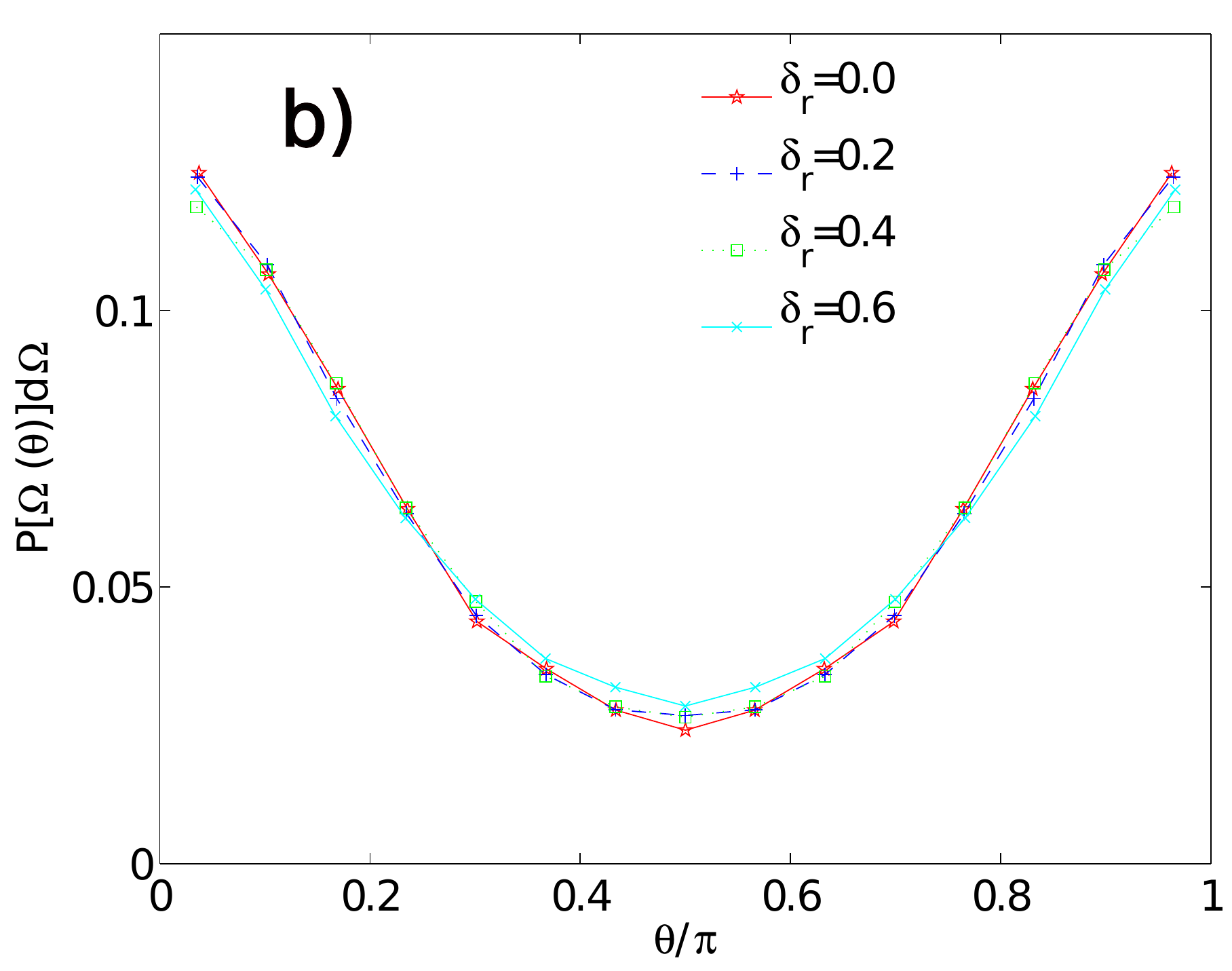}
\end{center}
\caption{\protect
        The anisotropy of the contact normal.
        {\bf (a)} Histogram of the angle $\theta$
            between the contact vectors and the positive $z$-semiaxis 
            for different shape parameters $\eta$ and $\zeta$, for the 
            mono-sized case ($\delta_r=0$).
        {\bf (b)} Histogram of the angle $\theta$ for 
            for a given shapen ($\eta=1.6$, $\zeta=1.8$) but different 
		size polidispersity. 
        Here $N\sim 3000-4500$ depending on the amount of the size polydispersity.
		}
\label{fig:contact-hist}
\end{figure}
\begin{figure}[htb]
\begin{center}
\includegraphics[width=0.45\textwidth,angle=0]{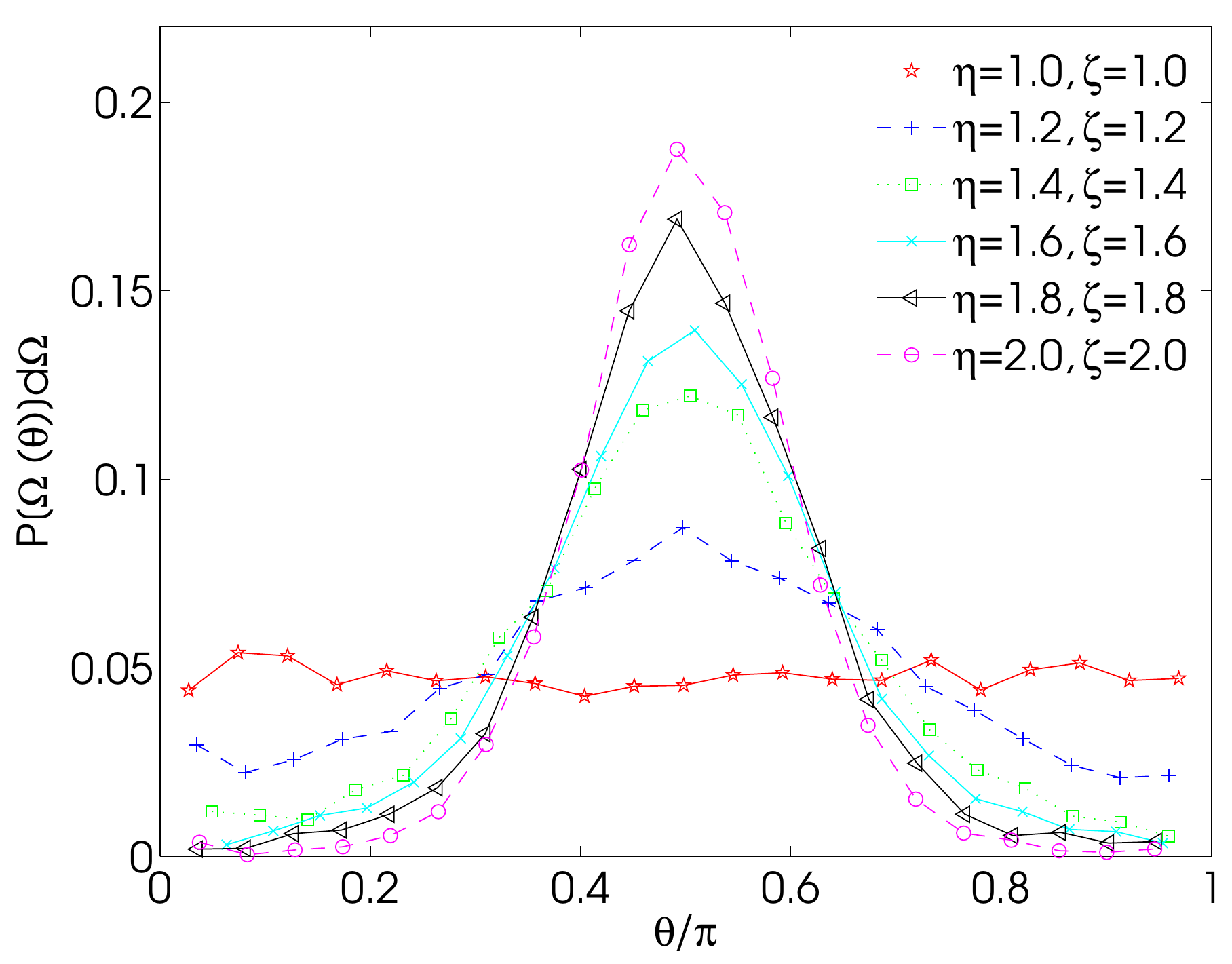}
\end{center}
\caption{\protect
     Distributions of the orientation angle $\theta$ between the
     largest semi-axis $a$ of each particle and the positive $z$-semi-axis.
     Each curves corresponds to particular values of the shape
     parameters.  Here $N\sim 3000$. }
\label{fig:orient-hist}
\end{figure}

Figure \ref{fig:contact-hist}a shows the results for deposits for
different values of the shape parameters $\eta$, $\zeta$ 
where $\delta_r=0$ . 
The vertical axis, $P(\Omega(\theta))d\Omega$, represents the fraction of 
number of vectors within the solid angle swept by a conic shell whose axis 
is parallel to $z$ and its vertex angle is $2\theta$. 
Note that the curves are symmetric with respect to $\theta=\pi/2$
because for each contact normal of a particle there is a contact normal 
with opposite direction from the other particle to which it is in contact.

The results show that for deposits of spheres 
the distribution $P(\Omega(\theta))$ is independent of $\theta$,
and consequently the configuration of the contacts is isotropic and has no preferred direction. 
However, as the shape of the particles deviates from a sphere 
the number of side contacts between particles, 
i.e.~the number of contacts on the horizontal plane, 
decreases and the contact normals line up along the vertical direction.
This is seen in Fig.~\ref{fig:contact-hist}a which shows for each shape 
there is a minimum at $\theta=\pi/2$ which corresponds 
to the number of contacts on the $(x,y)$-plane,
while having a maximum around $\theta=0$ 
(vertical direction). 
The minimum at $\theta=\pi/2$ together with the maximum at $\theta=0$
get more pronounced when increasing $\eta$ and $\zeta$. This means that for larger values of
$\eta$ and $\zeta$ which correspond to higher asphericities the particles  
tend to lie horizontally, for minimizing potential energy during deposition. 
Consequently most of the contacts will occur with the particles beneath and
above.  This can have dramatic effect on the force chains and the response of under 
shear and compression. Such a study is, however, beyond the scope the current work.

Figure~\ref{fig:contact-hist}a also shows
two crossing points at a value slightly higher than $\theta\sim\pi/5$, indicating
this direction as an invariant directions while varying the shape of
the depositing particles.

Figure \ref{fig:contact-hist}b shows the comparison of the angular 
distribution of contact normals for deposits with $\eta=6,\zeta=1.8$, 
with different polydispersities $\delta_r$. It is clear that the polydispersity has a minor effect. 

To study the anisotropy of the orientation of the particles we do a similar 
analysis for the principal directions of each ellipsoid. 
Figure \ref{fig:orient-hist} shows the results for the largest semi-axis 
$\vec{a}$ for the deposits.
One can see that as the shape of the particles deviates from a sphere the 
largest semi-axis $\vec{a}$ also tends to lie on the $(x,y)$-plane.
An invariant orientation is also observed, now for
$\theta\sim \pi/3$.
The polydispersities in the sizes of the particles 
has again no significant effects on the orientation anisotropy (not shown).

\section{Conclusions}
\label{sec:conclusions}

In this work, we studied systematically the static properties of 
deposits composed of ellipsoidal particles of different size and shape,
focusing on their density and anisotropy.
To this end, we developed an efficient Molecular Dynamics method for 
simulation of ellipsoidal particles which we used to generate deposits of 
such particles under gravity. 

For mono-sized particles, the density increases rapidly 
as the shape of the particles deviates from a 
sphere, reaching for particular values of the shape parameters a maximum 
significantly higher than the one observed in random close packing of spheres. 
With increasing further the shape parameters, that is, deviating further for a sphere 
density to decrease slowly.
The densities observed in the deposits are generally 
lower than the ones obtained through procedures for optimal packings.
Thus, as expected, deposition is not an ideal process for achieving 
the largest densities. 
Typically, porous media and media formed through deposition processes
present also densities much lower than the optimal one.
Such observations point toward the possibility for using deposition of
ellipsoids as a proper procedure for modeling porous media.
This is the subject of future works.

In the case of polydisperse particles the density takes higher values with 
qualitatively similar behavior.
A rather unexpected result is that the polydispersity in the sizes of the 
particles has almost no effect on the anisotropic behavior of the system.

We also study the anisotropy of such deposits and show that as the shape of 
the particles deviates from a sphere stronger anisotropies are observed. Such 
anisotropies appear due to the gravity and are absent in the systems without gravity.





\section*{Acknowledgements}

The authors thank Bibhu Biswal and Jens Harting for useful discussions
and the {\it Funda\c{c}\~ao para a Ci\^encia e a Tecnologia} for 
financial support, under the fellowship  {\it SFRH/BPD/48974/2008} and
contract {\it Ci\^encia 2007}.


\end{document}